# Deconstructing the Hubbard Hamiltonian by Ultrafast Quantum Modulation Spectroscopy in Solid-state Mott Insulators

S. Kaiser[1*], S.R. Clark[2,3 *], D. Nicoletti[1*], G. Cotugno[1,3], R.I. Tobey[3], N. Dean[3], S. Lupi[4], H. Okamoto[6], T. Hasegawa[5], D. Jaksch[3,2] and A. Cavalleri[1,3]

[1] *Max Planck Department for Structural Dynamics, Universität Hamburg, Germany*
[2] *Centre for Quantum Technologies, National University of Singapore, Singapore*
[3] *Department of Physics, Oxford University, Clarendon Laboratory, Parks Road, Oxford, UK*
[4] *IOR-CNR, Department of Physics, University of Rome "La Sapienza", Rome, Italy*
[5] *National Institute of Advanced Industrial Science and Technology, Tsukuba, Japan*
[6] *Department of Advanced Materials Science, University of Tokyo, Chiba 277-8561, Japan*

**Most available theories for correlated electron transport are based on the Hubbard Hamiltonian. In this effective theory, renormalized hopping and interaction parameters only implicitly incorporate the coupling of correlated charge carriers to microscopic degrees of freedom. Unfortunately, no spectroscopy can individually probe such renormalizations, limiting the applicability of Hubbard models. We show here that the role of each individual degree of freedom can be made explicit by using a new experimental technique, which we term "quantum modulation spectroscopy" and we demonstrate here in the one-dimensional Mott insulator ET-$F_2$TCNQ. We explore the role on the charge hopping of two localized molecular modes, which we drive with a mid infrared optical pulse. Sidebands appear in the modulated optical spectrum, and their linshape is fitted with a model based on the dynamic Hubbard Hamiltonian. A striking asymmetry between the renormalization of doublons and holons is revealed. The concept of quantum modulation spectroscopy can be used to systematically deconstruct Hubbard Hamiltonians in many materials, exposing the role of any mode, electronic or magnetic, that can be driven to large amplitude with a light field.**

Hubbard models[i,ii] are believed to capture the essential features of strongly-correlated-electron transport[iii] and have been widely applied to polymers, transition metal oxides, fullerenes and cuprates. However, such effective low-energy theories obscure the contributions of most microscopic degrees of freedom, which must be made explicit to describe correlated electron materials with predictive accuracy.

Here, we extend concepts routinely used in the study of strongly-correlated quantum gases[iv,v] to the solid state. We show that these couplings can be made experimentally explicit by optically modulating each microscopic degree of freedom separately and by measuring the perturbed optical spectrum. This new experimental technique is shown to be equivalent to a deconstruction of the Hubbard Hamiltonian, in which the modulated state exposes the nature of each microscopic coupling. As an example, we show that in the organic Mott insulator ET-$F_2$TCNQ a pronounced sideband manifold appears inside the Mott gap when one specific molecular mode is modulated. As quantitatively captured by a "locally-vibrating" dynamic Hubbard model[vi,vii], these lineshapes reveal asymmetry between the renormalization of holons and doublons. This allows investigating the different ways in which holes and electrons affect strongly correlated electron transport. That concept is of general interest for correlated materials notably in the superconducting cuprates.

The organic Mott insulator *bis*(ethylendithyo)-tetrathiafulvalene-difluorotetracyano-quinodimethane (ET-$F_2$TCNQ) is chosen as a prototypical system, in which model

Hamiltonian physics coexists with complex microscopic couplings. In this compound, chains of donor ET molecules form half-filled one-dimensional bands with strong onsite correlations and Mott-insulating properties[viii]. The static optical properties of ET-F$_2$TCNQ are reported in figure 1 for light polarized along the chain. A charge-transfer resonance, corresponding to excitations of the type (ET$^+$, ET$^+$) → (ET$^{2+}$, ET$^0$) is observed at photon energies ~ 700 meV (5500 cm$^{-1}$), reflecting the existence of a correlation gap.

Figure 1 shows how this excitation spectrum is well fitted by a model of the optical conductivity (see supplemental material) that starts from the $L$-site extended Hubbard Hamiltonian[ix],

$$\hat{H}_{\text{Hub}} = -t \sum_{\ell,\sigma} \left( c^\dagger_{\ell\sigma} c_{(\ell+1)\sigma} + \text{h.c.} \right) + U \sum_\ell \hat{n}_{\ell\uparrow} \hat{n}_{\ell\downarrow} + V \sum_\ell \hat{n}_\ell \hat{n}_{\ell+1}$$

with only the addition of a background contribution from a high frequency oscillator (dashed blue curve). In the expression for $\hat{H}_{\text{Hub}}$, $\hat{c}^\dagger_{\ell\sigma}$ $\hat{c}^\dagger_{l,\sigma}$ and $\hat{c}_{\ell\sigma}$ $\hat{c}_{l,\sigma}$ are creation and annihilation operators for an electron at site $\ell$ with spin $\sigma$, $\hat{n}_{\ell\sigma}$ $\hat{n}_{l,\sigma}$ is its corresponding number operator and $\hat{n}_\ell = \hat{n}_{\ell\downarrow} + \hat{n}_{\ell\uparrow}$.

Figure 2a visually clarifies the underlying microscopic physics of correlated charge-transfer excitations. Doubly occupied sites, hereafter referred to as "doublons", are created with repulsive energy $U$ and are bound to a neighbouring "holon" site with attractive energy $-V$.

As shown in figure 2b, the electronic configuration is strongly coupled to local modes. Typically, charge transfer "stiffens" the oscillator on the holon site and "slackens" it on the electron site[x,xi].

Theoretically the coupling to each degree of freedom can be made explicit by *deconstructing* the Hamiltonian

$$\widehat{H}_{\text{LVH}} = \widehat{H}'_{\text{Hub}} + \widehat{H}_{\text{I}} + \sum_{\ell} \widehat{H}_{\ell}$$

where $\widehat{H}'_{\text{Hub}}$ is renormalized by all the couplings except the one being analysed, leading to new effective $U'$, $V'$ and $t'$. The second term $\widehat{H}_{\text{I}}$ takes into account the coupling between the charge and the one local degree of freedom under scrutiny, with a type of coupling to be specified. The third term involves a sum over all lattice sites $\sum_{\ell} \widehat{H}_{\ell}$ accounting for the energy stored in a harmonic oscillator with angular frequency $\Omega$, displacement $\hat{q}_l$ and ground state size $a_0$.

In the ground state, the relative role of the last two terms $\widehat{H}_{\text{I}} + \sum_{\ell} \widehat{H}_{\ell}$ is vanishingly small. However, out of equilibrium, for example when the mode is non-thermally populated by coherent excitation with an optical pulse $\widehat{H}_{\text{I}} + \sum_{\ell} \widehat{H}_{\ell}$ becomes quantitatively important if the mode is non-thermally populated, for example by coherent excitation with an optical pulse. With suitable assumptions on the structure of $\widehat{H}_{\text{I}}$, key information can be revealed on the nature of the coupling by analysing the electronic spectrum perturbed in this way.

As a demonstration experiment here we expose the specific coupling of a charge-sensitive mode in ET-F$_2$TCNQ. We measured the time dependent changes of the optical properties as one localized intramolecular mode was driven with mid-infrared femtosecond pulses. These pulses were generated by parametric frequency conversion of the 800 nm-wavelength output of an amplified femtosecond laser. They were also polarized perpendicular to the chains, to minimize direct coupling with the electronic properties being probed. A single asymmetric, infrared active mode close to ~10 µm wavelength (1000 cm$^{-1}$) was excited[xii].

The time dependent optical properties of the solid were simultaneously probed in reflection over a broad spectral range as a function of pump-probe time delay. The tuneable output of an optical parametric amplifier was used to probe the reflectivity changes along the ET molecule chains in the mid (1800-3000 cm$^{-1}$) and near infrared (4000-7000 cm$^{-1}$). In the THz range (25-85 cm$^{-1}$) single-cycle pulses generated by optical rectification in ZnTe were used.

The 10-µm mode was excited with a maximum fluence of ~35 mJ cm$^{-2}$, corresponding to field strengths of approximately 10 MV/cm, and expected to strongly deform the molecular oscillator. A photo-induced red shift of the charge transfer band was observed, as well as new distinct peaks inside the gap (figure 3a). The relaxation back to the ground state occurs with a double-exponential decay ($\tau_1$ = 230 fs and $\tau_2$ = 4.5 ps), likely dictated by the lifetime of the hot vibrational mode.

In the driven state, no metallic response was detected in the THz range, where the reflectivity remained low and the phonon resonances unscreened (see figure 3b). This observation shows that the physics of vibrational excitation is different from that observed for above gap electronic excitations, where a Drude-like metallic response is typically observed[xiii,xiv]. Also, the absence of a metallic feature in the spectrum indicates that the observed effect is different from previous cases in which collective phonon excitations could drive phase changes[xv,xvi,xvii,xviii,xix].

Quantitative analysis of these results starts from fitting the broadband reflectivity spectra of figure 3b with a Drude-Lorentz Model and transforming it into optical conductivity[xx]. The conductivity lineouts are shown in figure 4c. The red-shift of the charge transfer band occurs from its equilibrium position at $\omega/c \approx 5500$ cm$^{-1}$ toward 5000 cm$^{-1}$, and a new band is observed approximately at 4200 cm$^{-1}$. Additionally, a mid-gap resonance and a weaker peak appeared at ~3000 cm$^{-1}$ and ~2000 cm$^{-1}$, respectively.

To analyse the data we apply a minimal model based on the idea of deconstructing the Hubbard Hamiltonian starting from the locally vibrating model Hamiltonian $\widehat{H}_{\text{LVH}}$ discussed above. From this the key contribution exposed is the coupling term $\widehat{H}_{\text{I}}$, which is assumed to be primarily coupled to the electronic configuration as $\hat{n}_\ell f(\hat{q}_\ell) + \hat{n}_{\ell\uparrow}\hat{n}_{\ell\downarrow}g(\hat{q}_\ell)$, with general functions $f$ and $g$ (see supplemental material). This goes substantially beyond typical electron-phonon interactions within the Holstein[xxi] model where only a linear coupling to the charge density of the form $\hat{n}_\ell \hat{q}_\ell$ is retained and is necessary in order to capture the experimental findings.

Within our model infrared active distortions cause the onsite wave function to oscillate with a "sloshing" motion, leading to a dipolar coupling to the charge sector (see supplemental material). Then to lowest order

$$\hat{H}_\text{I} = \sum_\ell (h_\ell \mathbb{H}_\ell - d_\ell \mathbb{D}_\ell) \hat{q}_\ell^2,$$

where $\mathbb{D}$ and $\mathbb{H}$ are projectors onto the doublon and holon states with respective positive couplings $d$ and $h$. The quadratic coupling causes the oscillators to stiffen to $\Omega_h$ when a holon is present and to slacken to $\Omega_d$ in the presence of a doublon (see figure 2b).

The real part of the optical conductivity $\sigma(\omega)$ of $\hat{H}_\text{LVH}$ was computed via the unequal time current-current correlation function[xxii] $\chi_{jj}(\omega)$. The narrow hopping bandwidth of ET-F$_2$TCNQ justifies $\sigma(\omega)$ being calculated analytically in the *atomic limit* where electron hopping is neglected. Intuitively, one can think of the vibrational excitation from an infinite mass molecular oscillator with a time $\tau$ dependent displacement $q = Q\cos(\Omega\tau)$. In this limit the onsite interaction coefficient is modulated by $(h - d)q^2$. Consequently a shift of the charge transfer resonance, by an amount that depends on the amplitude of the driven mode, and classical sidebands at multiples of $\pm 2\Omega$ on each side of the charge transfer resonance are predicted. The charge transfer resonance is expected to shift to the red if the renormalization around the doublon exceeds that of the holon, and to the blue for the opposite case.

Thus, the observed *red shift* in figure 4a of the charge transfer resonance carries an important qualitative insight of our experiment, i.e. that the renormalization of the molecular environment is far larger on the doublons than on the holons.

To reproduce the actual experimental line shape and estimate the coupling strength the oscillation requires a quantum mechanically treatment, allowing for a finite mass, and assuming strong coupling. Figure 4b and 4c show the comparison with the optical conductivity extracted from the reflectivity data and a calculation of $\sigma(\omega)$ as discussed above. We obtain a driving strength $Q/a_0 \approx 2$, and find that the doublon oscillator suffers a significant frequency reduction to $\Omega_d \approx 0.26\Omega$, while the holon's $\Omega_h \approx 1.10\Omega$ is only marginally increased. Therefore the major features seen in figure 4b are entirely a consequence of the slackening of the doublon oscillator as schematically shown in figure 4a. The reduced spacing between energy levels causes the transition frequencies to move into the gap, an effect that is significant only if the vibrational mode is appreciably populated. The remanents of the classical sidebands are located only at low frequencies and are now split into multiples of $\Omega - \Omega_d$. We note that the 10% modulation of the holon frequency is in agreement with the expectations[x,xi] for a thermal charge ordered state. However, the substantial modulation by a factor 4 on the doublon site goes far beyond any equilibrium predictions. Instead it indicated that large distortions were induced by strong driving to which a slackened mode is more susceptible. As the excited mode relaxes the distortion is reduced and the effects vanish as seen in figure 4c. At 0.5 ps after the excitation the peaks inside the gap are strongly reduced and the charge transfer band shifts back to its equilibrium position.

This response is mode selective. When the excitation wavelength was tuned away from the frequency of this vibration, the mid-infrared resonances disappeared. By way of comparison, we show what is observed when tuning the pump wavelength to 6 μm, near a weakly infrared-active symmetric mode along the chains[xxiii]. As shown in figure 5, only a reduction in spectral weight at the charge transfer resonance was observed, without significant response at other wavelengths. This is consistent with the weak linear coupling expected for this mode.

By using selective vibrational excitations we have deconstructed the Hubbard Hamiltonian, showing how individual microscopic excitations couple to the charge sector. In the organic compound ET-F2TCNQ, which has been analysed experimentally and theoretically, the charge transfer process creates highly asymmetric holon-doublon pairs due to coupling to one molecular vibration. One can speculate that as a general feature, double occupancies may be subjected to stronger renormalization and thus localisation, a phenomenon that may be related also to the higher effectiveness of hole doping in inducing High-Temperature superconductivity.

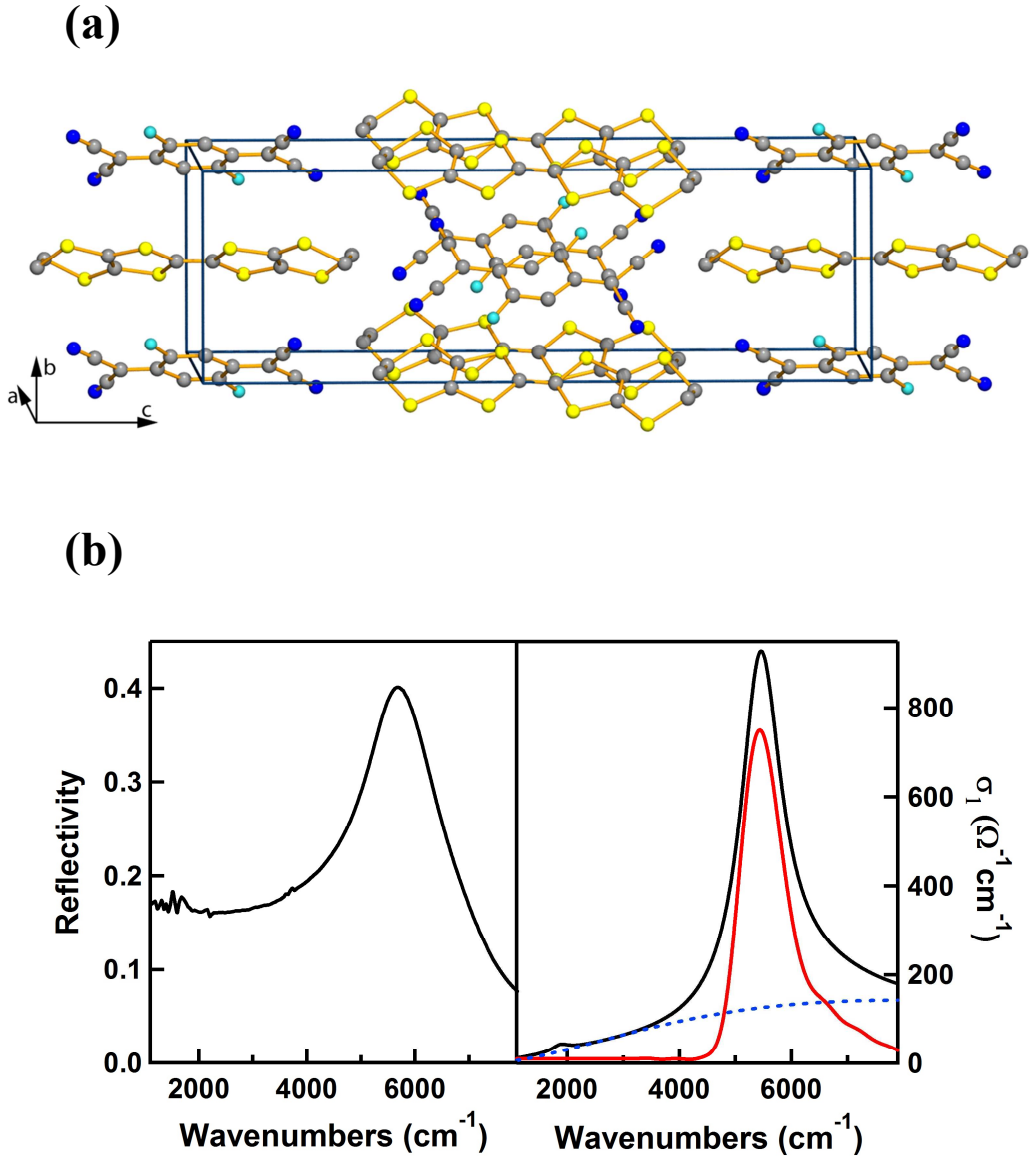

**Fig. 1 (a)** Crystal structure of the ET-F$_2$TCNQ. The ET molecules form a one-dimensional, half filled Mott-insulating chain along the a axis. **(b)** Black curves: static reflectivity (left panel) and optical conductivity (right panel) of ET-F$_2$TCNQ for light polarized along the a-axis. The optical conductivity is extracted using a Drude Lorentz fit of the reflectivity. The charge transfer peak in the optical conductivity is fitted to a ten site numerical solution of the Hubbard model (red) based on the two-time current-current correlation function (see supplemental material). The best fit is obtained with hopping $t \approx 50$ meV, onsite interaction $U \approx 800$ meV and nearest-neighbour repulsion $V \approx 100$ meV. These values are in agreement with reference ix and the size of the optical gap in reference viii. A quantitative match with the experimental conductivity is achieved by adding a Lorentzian oscillator centred at 8000 cm$^{-1}$, which takes into account contributions by higher-lying transitions (blue dashed curve).

**(a)**

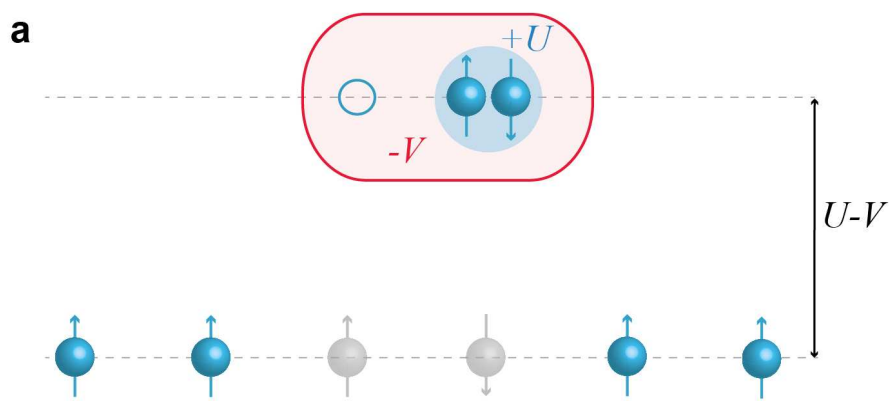

**(b)**

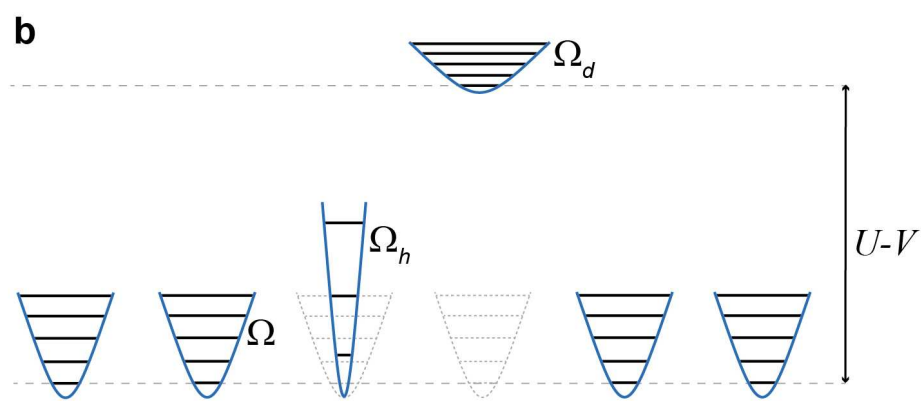

**Fig. 2 (a)** Each charge transfer excitations create an empty site (holon) and a doubly occupied site (doublon). The holon-doublon pair involves a repulsive energy $U$ and a binding energy of $V$, totalling $U - V$. **(b)** Electrons in the valence orbital of each ET molecule are coupled to many local degrees of freedom, to be thought of as a set of $L$ harmonic oscillators, one on each lattice site. Vibrational modes are expected to "stiffen" when one electron is removed, and to "slacken" when one electron is added.

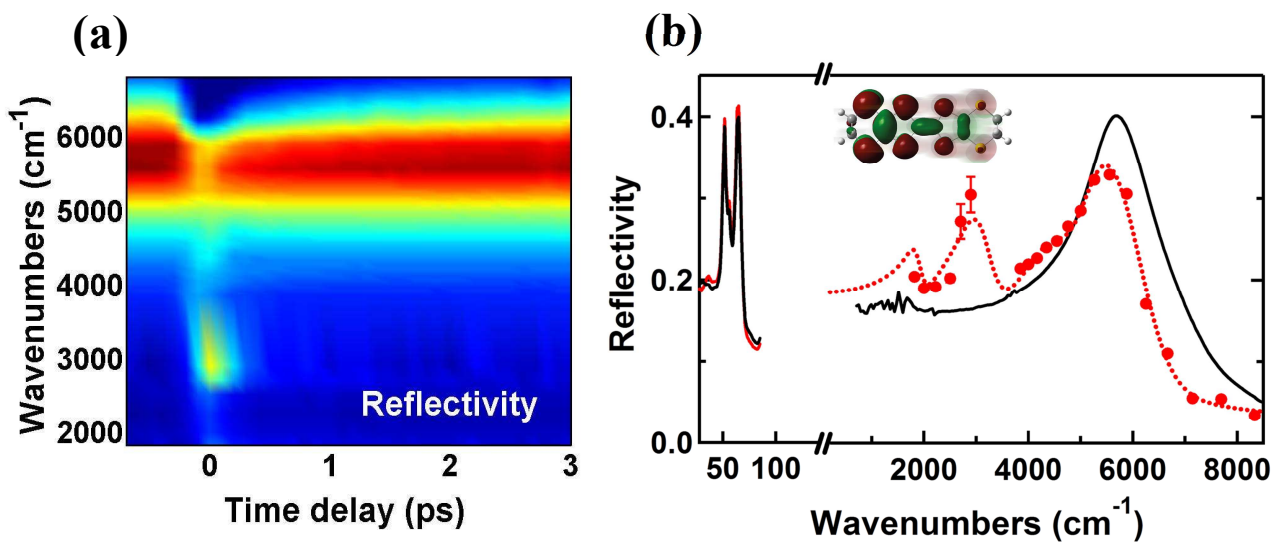

**Fig. 3 (a)** Time and frequency dependent reflectivity after selective modulation of the infrared active molecular mode at 10-μm wavelength. **(b)** Frequency dependent reflectivity at equilibrium (black) and at the peak of the modulating infrared field (red). Full circles indicate experimental data, whilst the dashed line is a Lorentz fit to the data to extract the optical conductivity. In the low-frequency range (below 100 cm$^{-1}$), the full lines indicate the equilibrium (black) and transient (red) reflectivity measured with single-cycle THz-pulses. The two peaks are phonon modes of the molecular crystal. In the inset we report a caricature of the "sloshing" distortions associated with the 10-μm mode, as calculated using Gaussian 03, assuming a frozen distortion along the normal mode coordinate.

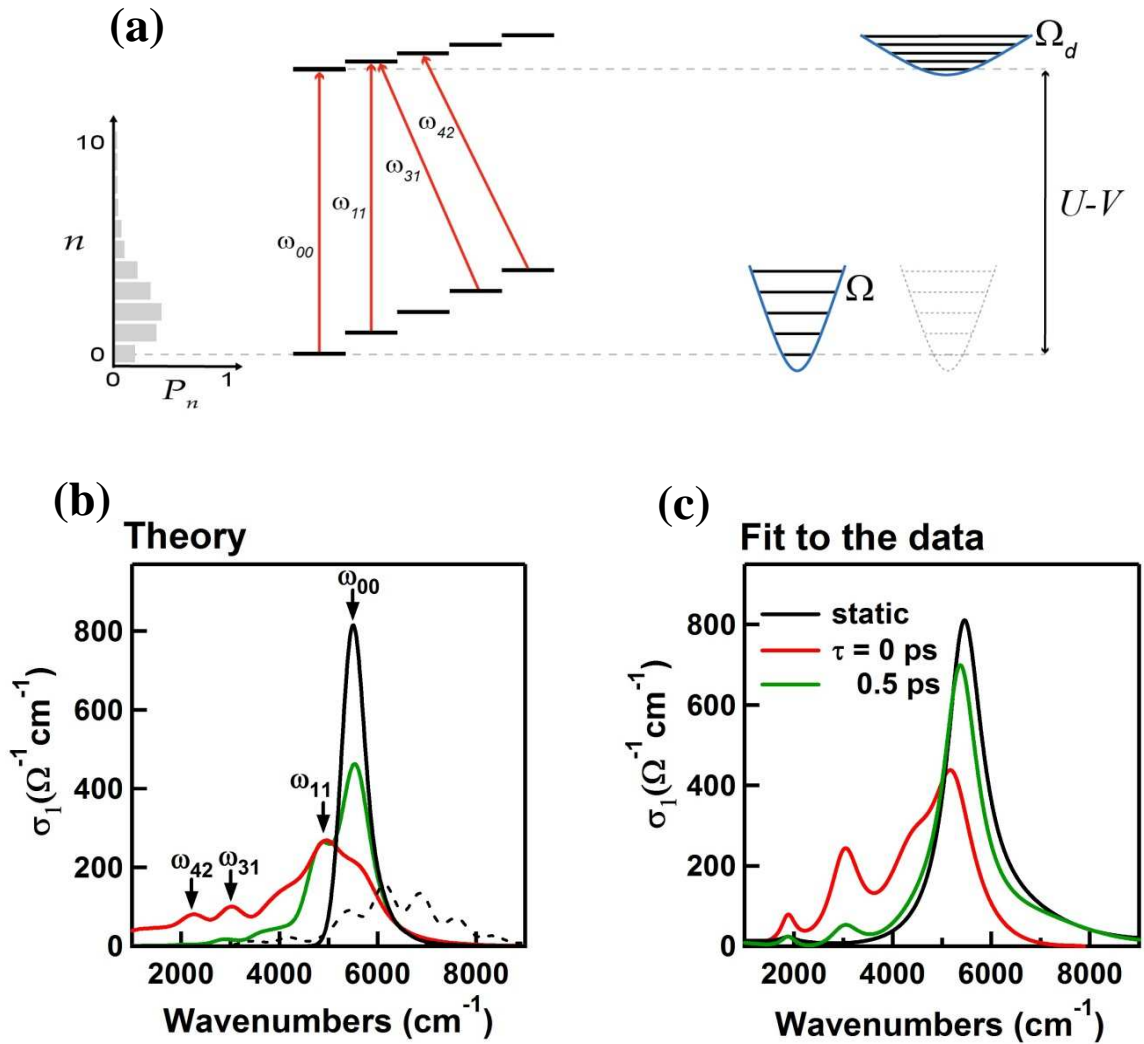

**Fig. 4 (a)** Schematic representation of the vibrational transitions assuming for simplicity $\Omega_h = \Omega$. The double-occupied site has a significantly renormalized vibration, which results in new transitions in the optical conductivity. The $p_n$ are the occupations of the driven lower vibrational ladder. **(b)** Theoretical optical conductivity at equilibrium (as in figure 1b) upon coherent excitation of the local vibrational mode. The green and red curve represent two different driving amplitudes. The various peaks are labelled according to the transitions of 4a. The dashed curve represents the case in which the asymmetry between holons and doublons are inverted, i.e. in which the holons are renormalized and doublons remain unperturbed. The dashed curve represents the case in which the asymmetry between holons and doublons are inverted, i.e. in which the holons are renormalized and the doublons remain unpertubed. **(c)** Optical conductivity calculated via Drude Lorentz fit to the reflectivity spectra of figure 3b, at two selected pump-probe delays. The background oscillator of figure 1b has been subtracted.

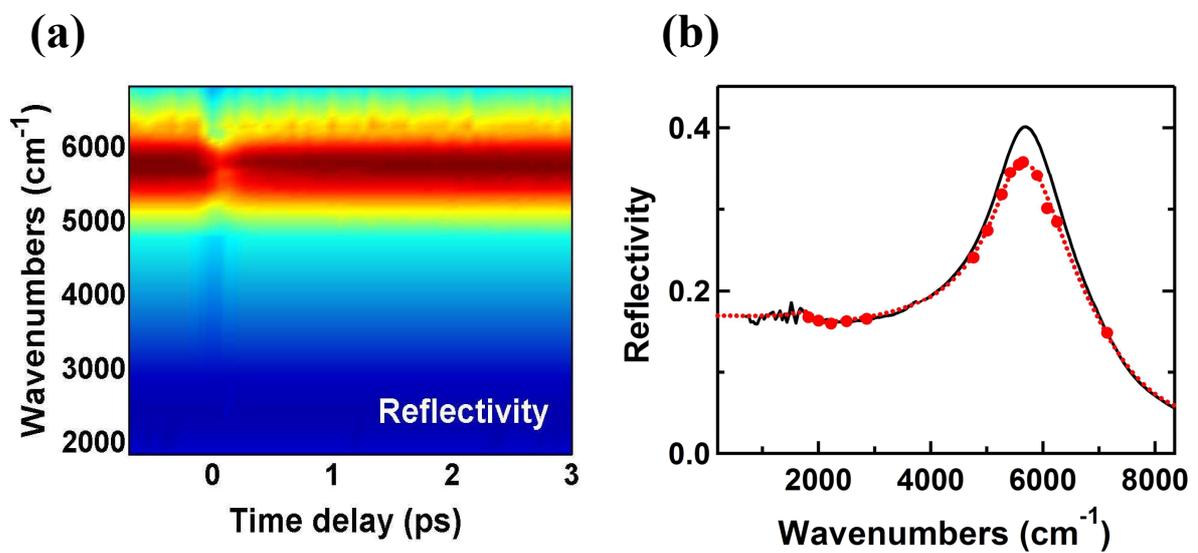

**Fig. 5 (a)** Time and frequency dependent reflectivity after modulation at 6 μm wavelength. **(b)** Frequency dependent reflectivity at equilibrium (black) and at the peak of the modulating infrared field (red). Full circles indicate experimental data, whilst the dashed line is a Lorentz fit to the data. .

**SUPPLEMENTAL MATERIAL**

**Coupling of local modes to electrons**

A general spin-independent coupling between a local harmonic oscillator and the electronic configuration can be written in terms of $\hat{n}_\ell$ and $\hat{n}_{\ell\uparrow}\hat{n}_{\ell\downarrow}$ only, as the operator $\hat{n}_\ell$ is fermionic. We assume that the coupling can be expressed as $\hat{n}_\ell f(\hat{q}_\ell) + \hat{n}_{\ell\uparrow}\hat{n}_{\ell\downarrow}g(\hat{q}_\ell)$, where $f(\hat{q}_\ell)$ and $g(\hat{q}_\ell)$ are two functions of the local mode coordinate that are not known a priori. By expanding the functions $f$ and $g$ into a Taylor series we obtain

$$\hat{H}_\text{I} = \sum_\ell \hat{n}_\ell (A_1 \hat{q}_\ell + A_2 \hat{q}_\ell^2 + \cdots) + \hat{n}_{\ell\uparrow}\hat{n}_{\ell\downarrow}(B_1 \hat{q}_\ell + B_2 \hat{q}_\ell^2 + \cdots)$$

with coupling constants $A_i, B_i$ which are constrained by the symmetry of the molecular modes[xxiv]. An antisymmetric infrared vibration, such as the 10-µm mode in the experiment, is akin to an oscillating dipole. Within the Born-Oppenheimer approximation this perturbation causes admixing of the valence orbital with higher-lying excited states of differing parity and induces an energy shift that is an even function of $\hat{q}_\ell$, meaning e.g. that $A_1 = 0$. A symmetric Raman vibration, such as the mode at 6 µm, is instead captured by an oscillating quadrupole, admixing higher-lying states of the same parity and causing a linear energy shift with non-zero $A_1$.

The second term in the expression for $\hat{H}_\text{I}$, which includes coupling to the double occupancy, is determined by computing the Coulomb repulsion arising from both electrons occupying the admixed vibrational orbital $|\psi_q\rangle$ via $U(q) \propto \langle\psi_q|_1\langle\psi_q|_2 \, \hat{r}_{12}^{-1} \, |\psi_q\rangle_1|\psi_q\rangle_2$ where $\hat{r}_{12} = |\hat{\boldsymbol{r}}_1 - \hat{\boldsymbol{r}}_2|$ is the distance between the electrons.

For the infrared vibration the differing parity of the states in the admixture causes $U(q)$ to again vary, to lowest order, quadratically with displacement $q$ (implying $B_1 = 0$), while the Raman vibration retains a linear dependence. It is found that $B_2 < 0$ for the infrared mode because the valence orbital spatially expands for any non-zero displacement in this simple model. We finally re-arrange terms to isolate the holon $\mathbb{H}_\ell = (1 - \hat{n}_{\ell\uparrow})(1 - \hat{n}_{\ell\downarrow})$ and doublon $\mathbb{D}_\ell = \hat{n}_{\ell\uparrow}\hat{n}_{\ell\downarrow}$ couplings and to give the expressions in the main text.

**Optical conductivity of a vibrating Hubbard chain**

Since $U \gg k_B T \gg t^2/U$ the system does not display any long-range Neel spin-order and does not contain doublons in its thermal state. It is thus well approximated by a half-filled completely spin-mixed state

$$\hat{\rho}_s = \frac{1}{2^L} \sum_{\vec{\sigma}} |\vec{\sigma}\rangle\langle\vec{\sigma}|$$

where $|\vec{\sigma}\rangle$ is a spin configuration state of the electronic Mott insulator. Given that $k_B T < \hbar\Omega$ higher vibrational states of the localised molecular oscillators are essentially unoccupied at room temperature. To model the vibrationally driven state we assume that each oscillator is prepared instantaneously in a coherent state $|\alpha\rangle$. Because the experiment does not control the phase of the driven oscillator, nor the relative phase between them, we phase average over $\alpha$ placing each oscillator into state $\hat{\rho}_\alpha$. The overall system state is $\hat{\rho} = \hat{\rho}_s \otimes \prod_\ell \hat{\rho}_\alpha$.

To compute the optical conductivity we start from the unequal time current-current correlation function which is defined as $\chi_{jj}(\tau) = \Theta(\tau)\mathrm{tr}[\hat{\jmath}(\tau)\hat{\jmath}(0)\hat{\rho}]$, where $\Theta(\tau)$ is the Heaviside function and $\hat{\jmath}(\tau) = \exp(i\hat{H}\tau/\hbar)\,\hat{\jmath}\,\exp(-i\hat{H}\tau/\hbar)$ is the Heisenberg picture current operator. The regular finite frequency optical conductivity then follows as

$$\sigma_1(\omega > 0) \propto \frac{1}{\omega} \mathrm{Re}\{\chi_{jj}(\omega > 0) - \tilde{\chi}_{jj}(\omega > 0)\}$$

where $\chi_{jj}(\omega)$ and $\tilde{\chi}_{jj}(\omega)$ are the Fourier transforms of $\chi_{jj}(\tau)$ and its complex conjugate, respectively.

In the atomic limit we obtain the current-current correlation function[xxii]

$$\chi_{jj}(\omega) \propto \int_{-\infty}^{\infty}\int_{-\infty}^{\infty} G_0(\omega'')G_h(\omega' - \omega'')G_d(\omega - \omega')\mathrm{d}\omega'\,\mathrm{d}\omega''$$

as a convolution of the Hubbard shift

$$G_0(\omega) \propto \int_{-\infty}^{\infty} \Theta(\tau)e^{-i(U-V-\hbar\omega)\tau/\hbar}d\tau$$

and the onsite vibrational doublon and holon correlation functions

$$G_{h,d}(\omega) \propto \sum_{n,m} p_n |\langle n|\hat{S}(\xi_{h,d})|m\rangle|^2 \delta\left(\omega - \left(m+\frac{1}{2}\right)\Omega_{h,d} + \left(n+\frac{1}{2}\right)\Omega\right)$$

where $\hat{S}(\xi_{h,d})$ is the squeezing operator with $\xi_{h,d} = \log\sqrt{\Omega_{h,d}/\Omega}$ and $p_n$ are the diagonal elements of $\hat{\rho}_\alpha$. The matrix elements $|\langle n|\hat{S}(\xi_{h,d})|m\rangle|^2$ weighting the $\delta$ functions are Franck-Condon factors and are well known analytically[xxv]. A similar analysis can proceed for the Raman mode.

**Atomic limit and hopping**

The optical conductivity is calculated here in the so-called zero-bandwidth atomic limit, where the ratio $\sigma_1(\omega)/t^2$ can be obtained exactly, which is justified by the parameters of ET-F$_2$TCNQ. For the bare Hubbard model the inclusion of hopping is known to cause broadening of the optical peaks[xxvi] on the order of $t$ and a shift of the order of $t^2/(U-V)$. In the dynamic Hubbard model $\hat{H}_\text{LVH}$, both linear and quadratic electron-oscillator coupling cause a suppression of $t$, analogous to polaronic effects[xxi]. Further suppression of the coherent hopping processes occurs once the vibrational modes are driven, similar to the reduction with temperature seen for polarons. For the strong coupling infrared case we estimate that the hopping of holons drops to around $0.2t$, while for doublons it is just $0.05t$. Moreover, because the oscillator coupling here is to a *local* molecular mode on each site, as opposed to a *collective* bath of lattice phonons, vibrational excitation disorders hopping through the chain further inhibiting motion. Thus, in combination with vibrationally activated incoherent hopping, the effects of a finite $t$ are to broaden the dominant contribution to the optical conductivity already captured by the onsite vibrational dynamics. To account for these mechanisms, as well as the spectral limitations of the measurement itself, we introduced an artificial broadening of $0.5t$ to the results presented.